\documentclass[a4paper,11pt]{article}
\pdfoutput=1 

\usepackage{jheppub_modified} 

\usepackage[T1]{fontenc} 

\usepackage{verbatim}
\usepackage{color}
\usepackage{graphicx}
\usepackage{subcaption}

\usepackage{sectsty}
\allsectionsfont{\boldmath}

\usepackage{xspace}
\usepackage[dvipsnames,table]{xcolor}
\usepackage{tabularx}
\usepackage{mathtools,amsmath,empheq,slashed}

\usepackage{bigcenter}

\newcolumntype{C}[1]{>{\centering\arraybackslash}p{#1}}

\newcommand{\eq}[1]{Eq.~(\ref{#1})}
\newcommand{\bib}[1]{Ref.~\cite{#1}}

\newcommand{\fig}[1]{Fig.~\ref{#1}}
\newcommand{\tab}[1]{Table~\ref{#1}}

\newcommand{\sect}[1]{Section~\ref{#1}}

\newcommand{\bea}{\begin{eqnarray}}
\newcommand{\eea}{\end{eqnarray}}

\newcommand{\eps}{\epsilon}

\newcommand{\crn}{\nonumber \\}

\newcommand{\fr}{\frac}

\def\MG5{{\tt MadGraph5\_aMC@NLO}}

\newcommand{\gev}{{\unskip\,\text{GeV}}}

\newcommand{\pb}{{\unskip\,\text{pb}}}

\newcommand{\U}{\mathcal{U}}
\newcommand{\oo}{\mathcal{O}}

\title{Unparticle effects at the MUonE experiment}

\author[a]{Duc Ninh Le,}
\author[b,c]{Van Dung Le,}
\author[d]{Duc Truyen Le,}
\author[b,c]{Van Cuong Le}

\affiliation[a]{Faculty of Fundamental Sciences, PHENIKAA University, Hanoi 12116, Vietnam}
\affiliation[b]{Department of Theoretical Physics, University of Science, Ho Chi Minh City 70000, Vietnam}
\affiliation[c]{Vietnam National University, Ho Chi Minh City 70000, Vietnam}
\affiliation[d]{Department of Physics, National Tsing Hua University, Hsinchu, Taiwan 300044, R.O.C.}

\emailAdd{ninh.leduc@phenikaa-uni.edu.vn}
\emailAdd{dunglvht@gmail.com}
\emailAdd{leductruyenphys@gapp.nthu.edu.tw}
\emailAdd{lvcuong.98tv@gmail.com}

\abstract{We investigate possible effects of unparticles at the MUonE experiment by considering a general model 
for unparticle with broken scale invariance, characterized by the scaling dimension $d$ and the energy scale $\mu$ at which the scale invariance is broken. Taking into account available relevant constraints on the couplings of the unparticles with the Standard Model (SM) leptons, we found that the MUonE experiment at the level of 10 ppm systematic accuracy is sensitive to such effects if $1<d\lesssim 1.4$ and $1\le \mu \lesssim 12$ GeV for vector unparticles. The effects of scalar unparticles are too feeble to be detected. The vector unparticles can induce a significant shift on the best-fit value of $a_\mu^\text{had}$ at the MUonE, thereby providing an opportunity to detect unparticles or to obtain 
a new bound on the unparticle-SM couplings in the case of no anomaly.}

\begin{document}
\maketitle
\flushbottom

\section{Introduction}
\label{sect:intro}
Unparticle is an interesting idea proposed by Georgi in 2007 \cite{Georgi:2007ek}. 
The idea is that there exists a new hidden sector which is scale invariant. 
This sector is called unparticle. Unparticle can interact with the Standard Model (SM) fields, leading to possible 
effects in precision low energy experiments.  

A scheme for unparticle can be sketched as follows. At a high energy regime above the SM scale, there exists 
a UV-completion theory with the SM fields and new fields called Banks-Zaks (BZ) fields \cite{Georgi:2007ek}. 
These two sets of fields interact via exchanging particles with masses of the order of $\Lambda_\text{UV}$ or higher. 
Below this energy scale, the effective interaction between the SM fields and the Banks-Zaks fields reads
\bea
\mathcal{L}_\text{UV} = \fr{c_\text{UV}}{\Lambda_\text{UV}^{d_\text{SM}+d_\text{UV}-4}} \mathcal{O}_\text{SM} \mathcal{O}_\text{UV},
\label{L_UV}
\eea
where $\mathcal{O}_\text{SM}$ is an operator with mass dimension $d_\text{SM}$ built out of the SM fields and
$\mathcal{O}_\text{UV}$ an operator with mass dimension $d_\text{UV}$ built out of the BZ fields, 
$c_\text{UV}$ is a dimensionless coupling.   
At some scale $\Lambda_\U < \Lambda_\text{UV}$, the interactions among the BZ fields then cause dimensional transmutation to 
form a scale-invariance system, named above as unparticle. The effective interaction between the SM fields and the unparticle 
is given by
\bea
\mathcal{L}_{\text{SM}-\U} = \fr{c_\U}{\Lambda_\U^{d_\text{SM}+d-4}} \mathcal{O}_\text{SM} \oo_\U,
\label{L_Unparticle}
\eea
where $c_\U$ is a dimensionless coupling, $d$ is the mass dimension of the unparticle operator. 

Unparticle can be thought of as a set of $d$ massless particles \cite{Georgi:2007ek}. 
The novel feature here is that $d$ is non-integral. This system is, by definition, scale invariant. 
It is important to note that we do not require unparticle to be conformal invariant, which is more restrictive 
(see e.g. \cite{Grinstein:2008qk} for the distinction between scale and conformal invariance). 
For example, for a conformal field theory, unitary constraint imposes a lower limit on the value of $d$, 
namely $d \ge 1$ for a scalar unparticle system and $d \ge 3$ for a vector case \cite{Mack:1975je,Grinstein:2008qk}. 
Such a constraint on the scaling dimension is not demanded in this study. As done in most studies in the literature, we will focus on the range $1<d<2$, which is most natural since it is close to the particle limit of $d=1$. Moreover, 
unparticle effects are largest in this region.

The original idea of Georgi \cite{Georgi:2007ek} assumes that unparticles exist below the scale $\Lambda_\U$. It was then very soon realized that data from cosmology and low-energy experiments puts severe limits on the couplings between the unparticles and the SM sector, see e.g. \cite{Davoudiasl:2007jr,Liao:2007bx,Balantekin:2007eg}, making it impossible to observe unparticle effects at present or near-future experiments. However, these constraints can be evaded if the scale invariance is broken at a energy scale $\mu$ which is sufficiently large compared to the scales of the cosmology and low-energy experiment processes. \bib{Barger:2008jt} found that $\mu \gtrsim 1\gev$ is enough. The case $\mu > M_Z$ was studied in \cite{Rizzo:2007xr}, and 
$1 \lesssim \mu < M_Z$ in \cite{Barger:2008jt}, using a simple model proposed in \cite{Fox:2007sy} for unparticle with broken scale invariance. In the limit $\mu \to 0$ the unbroken case of Georgi is then recovered. 

Recently, a new experiment named MUonE has been proposed with the letter of intent submitted to CERN in 2019 \cite{Abbiendi:2677471}. 
In this experiment, the differential cross section of the elastic $e\mu\to e\mu$ scattering, occurring at the 
energy of $0.4\gev$, will be measured with very high accuracy. 
All the systematics effects are expected to be known at 10 ppm \cite{Abbiendi:2677471}. 
If this experiment is realized it will be an excellent place to probe unparticles in the region of $\mu \approx 1\gev$.  
The purpose of this work is to study unparticle effects at the MUonE experiment taking into account the latest available constraints 
on the unparticle-SM couplings. 

The paper is outlined as follows. The unparticle models used in this work are first described in \sect{sect:unparticle}. 
In \sect{sect:constraint}, we then survey the available constraints on the unparticle-SM couplings using data provided in the literature. 
In this section, we provide a new bound on the (pseudo-)scalar unparticle couplings using the mono-photon cross section data at LEP2. 
This result will be useful for unparticle studies. Basics of the MUonE experiment are summarized in \sect{muone_exp}. 
Our main results are presented in \sect{unparticle_MUonE} where an analytical formula for calculating unparticle effects at the MUonE 
is given as well as numerical results for sensitivity curves and shifts on the best-fit value of $a_\mu^\text{had}$. 
Summary and conclusions are provided in \sect{sect:conclusion}. 
\section{Unparticle models}
\label{sect:unparticle}
In this work we are interested in unparticle effects at the MUonE experiment, the dominant contribution comes from 
the interactions between an unparticle and the charged leptons (electron or muon). We assume here lepton universality and no flavor-number violation for simplicity. The relevant SM operator in \eq{L_Unparticle} is therefore $\mathcal{O}_\text{SM} = \overline{f} \Gamma f$ with $\Gamma=$ $\mathbb{I}$, $\gamma_5$, $\gamma_\mu$, $\gamma_\mu \gamma_5$. We will consider these cases separately. We note that the unparticles can couple to 
other SM fields such as the quarks, the gauge bosons, and the Higgs boson. These effects are however much weaker, hence are here neglected.

To be specific, we consider the following four operators \cite{Georgi:2007ek,Georgi:2007si,Cheung:2007ap}:  
\begin{align}
\frac{c_S}{\Lambda_\U^{d-1}}\overline{f} f \oo_\U, \quad \frac{i c_P}{\Lambda_\U^{d-1}}\overline{f} \gamma_5 f \oo_\U, 
\quad \frac{c_V}{\Lambda_\U^{d-1}}\overline{f} \gamma_\mu f \oo_\U^\mu, 
\quad \frac{c_A}{\Lambda_\U^{d-1}}\overline{f} \gamma_\mu\gamma_5 f \oo_\U^\mu, 
\label{eq:models}
\end{align}
which are called scalar, pseudo-scalar, vector, and axial-vector unparticle models, respectively. 
The parameters $c_i$ ($i=S,P,V,A$) are dimensionless couplings, 
$\oo_\U$ is a scalar unparticle operator, $\oo^\mu_\U$ a vector unparticle operator. 
Since the value of $\Lambda_\U$ is unknown, we trade it to $M_Z$ by rewriting the couplings as follows
\begin{align}
c_X=\lambda_X\left(\frac{\Lambda_\U}{M_Z}\right)^{d-1},\;\; X=S,P,V,A.
\label{eqn:1}
\end{align}
The operators in \eq{eq:models} then read:
\begin{align}
\frac{\lambda_S}{M_Z^{d-1}}\overline{f} f \oo_\U,\quad 
\frac{i\lambda_P}{M_Z^{d-1}}\overline{f} \gamma_5 f \oo_\U,\quad
\frac{\lambda_V}{M_Z^{d-1}}\overline{f} \gamma_\mu f \oo_\U^\mu,\quad 
\frac{\lambda_A}{M_Z^{d-1}}\overline{f} \gamma_\mu\gamma_5 f \oo_\U^\mu.
\label{vertices_lambda}
\end{align}
The propagators of the scalar and vector unparticles have the following interesting forms \cite{Georgi:2007si,Grinstein:2008qk}
\begin{align}
\Delta_F(k)&=\fr{iZ_{d}}{(-k^2-i\eps)^{2-d}},\label{prop_scalar}\\
\Delta^{\mu \nu}_F(k)&=\fr{iZ_{d}}{(-k^2-i\eps)^{2-d}}(-g^{\mu\nu}+a\fr{k^\mu k^\nu}{k^2}),
\label{prop_vec}
\end{align}
where $k$ is the momentum of the unparticle,
\begin{align}
A_{d}=\frac{16\pi^{5/2}}{(2\pi)^{2d}}\frac{\Gamma(d+1/2)}{\Gamma(d-1)\Gamma(2d)}, \quad Z_{d} = \fr{A_d}{2\sin(d \pi)}.
\end{align}
The parameter $a$ is only relevant for the axial-vector case as its dependence is canceled out in the case 
of vector unparticle.  
Its value depends on the nature of the unparticle: $a=1$ if the operator 
$\oo_\U^\mu$ is transverse \cite{Georgi:2007si}, or $a = 2(d - 2)/(d - 1)$ for $d\ge 3$ if conformal invariance 
is required \cite{Grinstein:2008qk}. Since $1 < d < 2$ in this work, we will set $a=1$ in the numerical results. 
We have checked that the dependence on $a$ is completely negligible, i.e the change of the differential cross section in \eq{diff_XS_SMU} 
is less than $1$ ppt when varying $a\in [0,2]$.
\begin{figure}[h!]
  \centering
  \includegraphics[width=0.6\textwidth]{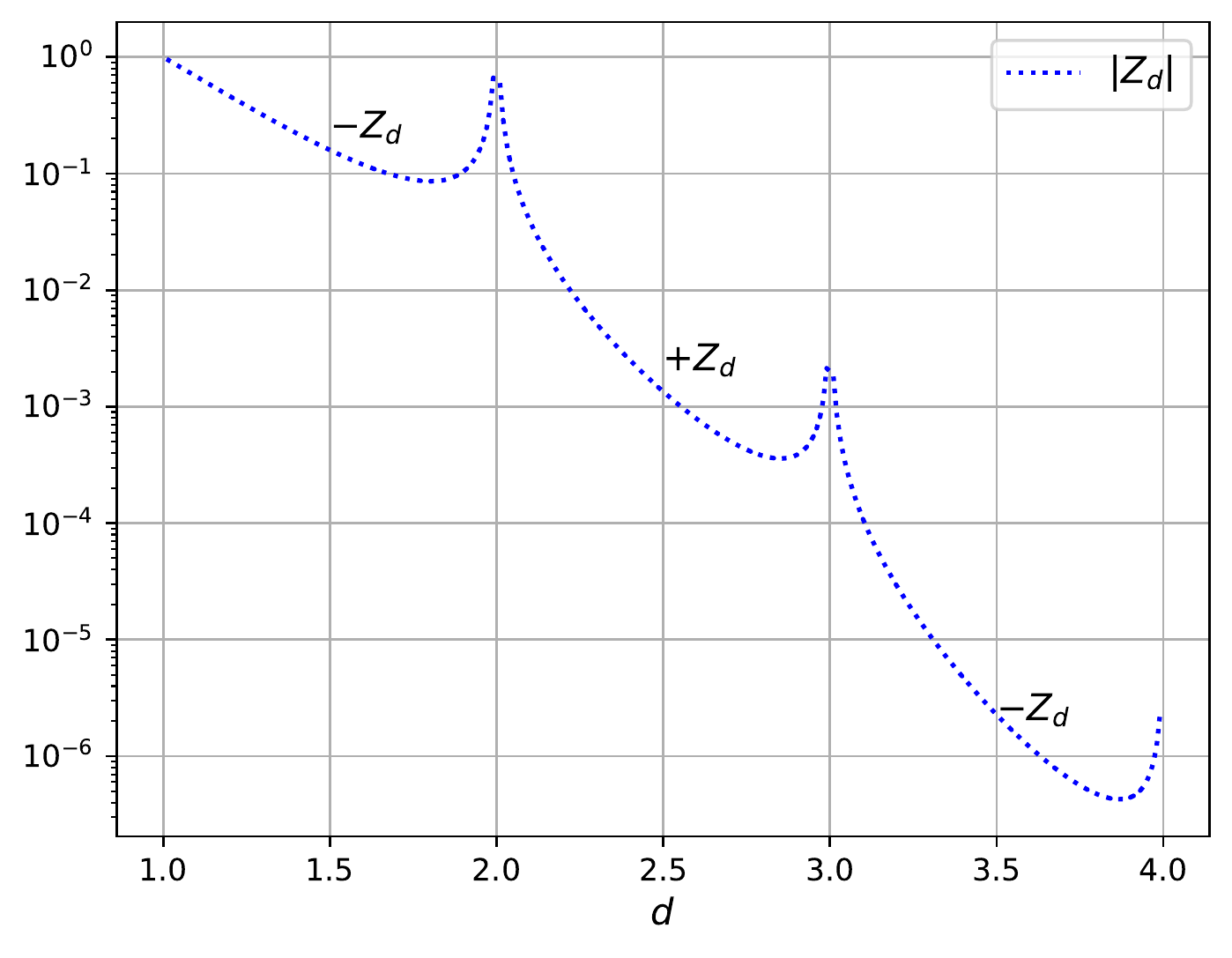}
  \caption{Absolute value of the $Z_d$ factor as a function of the scaling dimension $d$. 
  $Z_d$ is negative when $d$ is in $(1,2)$ or $(3,4)$, positive in $(2,3)$.}
  \label{fig:Zd}
\end{figure}

The factor $Z_{d}$ is plotted in \fig{fig:Zd}. 
Because of the denominator $\sin(d \pi)$ this factor is singular at integer values $d = n$ with $n\ge 2$ as shown in the plot.

We remark that the above unparticle propagators 
are very different from the usual particle ones. 
It is characterized only by the scale dimension $d$ 
of the unparticle operators. 
It is interesting to notice that the unparticle propagator (scalar or vector) reproduces 
the propagator of a massless particle in the limit $d \to 1$. 
Note that, if conformal invariance is required then $d\ge 3$ for the (axial-)vector cases \cite{Grinstein:2008qk}, thereby excluding the limit $d \to 1$.   
Since kinetic singularity occurs 
in the limit $E_\U \to 0$ (i.e. vanishing unparticle energy) in the case of $d < 1$ \cite{Georgi:2007ek}, we 
will focus on the range $1 < d < 2$ for both the scalar and vector unparticles, as discussed in the introduction.

The nice thing of unparticle is that it provides a new propagator, leading to novel interference effects with the SM amplitudes \cite{Georgi:2007si}. This may help us to explain new effects which will be discovered in the future.  

Adding broken scale invariance is relatively simple, as done in \cite{Fox:2007sy}. In this model, it is assumed that 
the contribution from the broken phase (i.e. energy less than $\mu$) is suppressed. The modes with energy less than 
$\mu$ are therefore removed from the spectral density function. This affects the unparticle propagator. 
The new propagators taking into account the effect of broken scale invariance now read \cite{Barger:2008jt}
\begin{align}
\Delta_F(k)&=\fr{iZ_{d}}{(-k^2+\mu^2-i\eps)^{2-d}},\label{prop_scalar_b}\\
\Delta^{\mu \nu}_F(k)&=\fr{iZ_{d}}{(-k^2+\mu^2-i\eps)^{2-d}}(-g^{\mu\nu}+a\fr{k^\mu k^\nu}{k^2}),
\label{prop_vec_b}
\end{align}
which are obtained from \eq{prop_scalar} and \eq{prop_vec} by the replacement $(k^2+i\eps) \to (k^2 - \mu^2 + i\eps)$. 
The full scale invariance case as originally proposed by Georgi is recovered in the limit $\mu \to 0$.
\section{Constraints on unparticle-SM interactions}
\label{sect:constraint}
Most studies on the literature have been focusing on the case of exact scale invariance, namely $\mu = 0$. 
In this limit, it was found that the Big Bang Nucleosynthesis (BBN) \cite{Davoudiasl:2007jr} and SN 1987A \cite{Davoudiasl:2007jr,Freitas:2007ip,Hannestad:2007ys,Das:2007nu,Dutta:2007tz} constraints put a severe 
limit on the strength of the unparticle-SM interactions, with the former giving a more stringent bound \cite{Davoudiasl:2007jr,Barger:2008jt}. For $1 < d < 2$, the coupling $|\lambda_V|$ between the vector unparticle and the SM fermions in \eq{vertices_lambda} must be smaller than $10^{-7}$ \cite{Davoudiasl:2007jr,Barger:2008jt}. Since this constraint was obtained using dimensional analysis \cite{Davoudiasl:2007jr}, and the dimensions of the vector and scalar unparticles in \eq{vertices_lambda} are exactly the same, we expect similar result for the (pseudo-)scalar and axial-vector unparticle couplings, namely $|\lambda_S|$, $|\lambda_P|$, $|\lambda_A|$ $< 10^{-7}$. With these tiny couplings, it is impossible to detect unparticle effects at the present or near-future collider experiments. 

For the case $\mu \gtrsim 1$~GeV, the BBN and SN 1987A constraints are evaded because 
the scale invariance is broken at the energy scale $\mu$ sufficiently large compared to the relevant energy scales, $1$~MeV for BBN and $30$~MeV for SN 1987A, as already observed in \cite{Barger:2008jt}. For the same reason, other constraints from low energy experiments such as the electron and muon anomalous magnetic moments \cite{Cheung:2007zza}, positronium decays \cite{Liao:2007bx}, neutrino decays into unparticles \cite{Anchordoqui:2007dp}, neutrino-electron scattering \cite{Balantekin:2007eg} are evaded as well.        

In this paper we consider the MUonE experiment where the scattering energy is $\sqrt{s}\approx 0.4\gev$ in the center-of-mass system (c.m.s), focusing on the region of the parameter space where $1 < d < 2$ and $\mu \ge 1\gev$. 
It is then clear that the unparticle effects are largest in the region of $d \approx 1$ and $\mu \approx 1\gev$ 
and decrease when $d$ or $\mu$ gets larger. We will show that the MUonE experiment at 10 ppm accuracy is insensitive 
to the region of $\mu > 12\gev$.     
\begin{figure}[h!]
  \centering
  \includegraphics[width=0.6\textwidth]{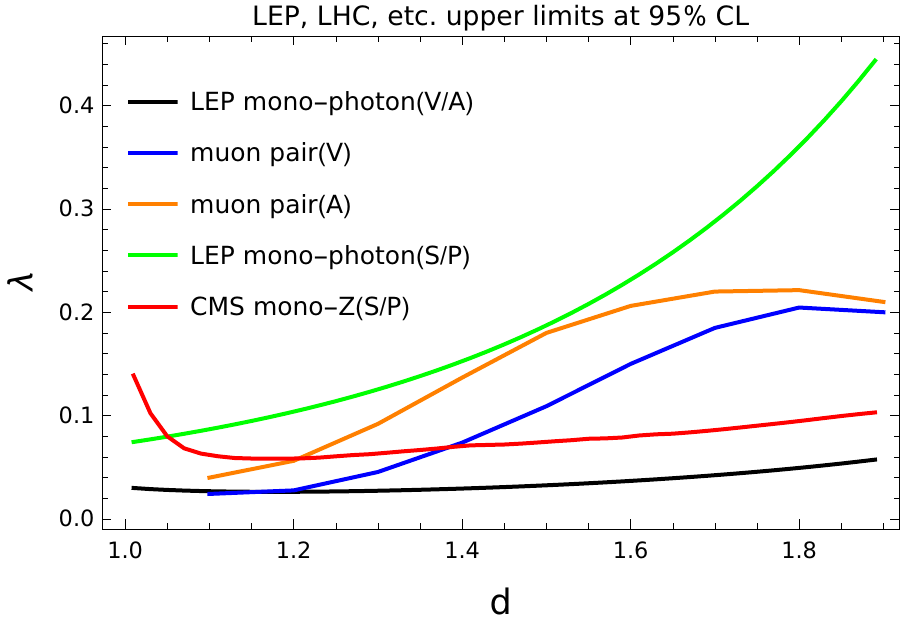}
  \caption{Upper limits at $95\%$ CL from LEP, CMS and other experiments (see text) data on (axial-)vector and (pseudo-)scalar unparticle parameters. The regions above the curves are excluded.}
  \label{bound_U}
\end{figure}

For the case $1 \le \mu \lesssim 12\gev$, we will take into account the available constraints from mono-photon production at LEP2 for (pseudo-)scalar and (axial-)vector unparticles \cite{L3:2003yon}, from $\mu^+\mu^-$ cross section and forward-backward assymmetry data at LEP-Aleph \cite{ALEPH:1999smx}, KEK-Venus \cite{VENUS:1997bjf,VENUS:1990tbk,VENUS:1990vwh}, PETRA-MarkJ \cite{MARKJ:1988agx,Mark-J:1985xqi} for (axial-)vector unparticles \cite{Barger:2008jt}, from mono-$Z$ production at the LHC-CMS for (pseudo-)scalar unparticles \cite{CMS:2020ulv}. These are the strongest and 
most relevant constraints on the unparticles that we can find in the literature. They are plotted in \fig{bound_U}. 
Bound from the mono-$Z$ production at LEP2 \cite{L3:1999ags} has been ignored because it is 
weaker compared to the mono-photon data \cite{Barger:2008jt}. 

The $95\%$ CL upper limit (black line) on the (axial-)vector unparticle couplings has been obtained using the differential cross section of the process $e^+ e^- \to \gamma + \text{unparticle}$ provided in \cite{Cheung:2007ap,Barger:2008jt}. 
For the case of (pseudo-)scalar unpartiles, our calculation gives
\begin{align}
d\sigma = \fr{A_d e^2\lambda_i^2 E_\gamma}{16\pi^3M_Z^2 s}\left(\fr{s-2\sqrt{s}E_\gamma-\mu^2}{M_Z^2}\right)^{d-2}\fr{\cos^2\theta_\gamma}{1-\cos^2\theta_\gamma}dE_\gamma d\Omega,\label{mono_gamma_S} 
\end{align}
where $\lambda_i = \lambda_S$, $\lambda_P$ and $\sqrt{s}$ is the c.m.s energy. 
Using the LEP2 $95\%$ CL upper limit of $\sigma \approx 0.2\pb$ obtained 
with the kinematic cuts $E_\gamma\in [5\gev,(s-\mu^2)/(2\sqrt{s})]$, $|\cos\theta_\gamma| < 0.97$, $\sqrt{s} = 207$ GeV 
from \bib{L3:2003yon} we then get the upper bound on the coupling $\lambda_i$ as plotted in \fig{bound_U} (green line). 
This new result will be useful for other unparticle studies.

In \fig{bound_U} we have set $\mu = 0$ for the mono-photon and mono-$Z$ constraints. 
For higher values of $\mu$, the unparticle cross sections get smaller, leading to higher upper limits on the couplings. 
However, as long as the value of $\mu$ remains small compared to the colliding energy, the $\mu$ dependence is tiny and can 
be safely neglected. The muon-pair bounds are taken from \cite{Barger:2008jt}, obtained with $\mu=1.5\gev$. 
Note that, the curves stop at $d=1.1$ because smaller values are not provided.  
These constraints are weaker than the mono-photon one, except in the region of $d < 1.2$ where 
the muon-pair bound on $\lambda_V$ is marginally more stringent.  
For the case of vector unparticles, the most relevant constraint is therefore the LEP2 mono-photon bound. 
For the scalar unparticles, the LEP2 mono-photon and CMS mono-$Z$ constraints are 
complementary. The former is more stringent when $d < 1.05$ and gets weaker with increasing $d$. 

\fig{bound_U} shows some interesting behaviors. The muon pair bounds are more stringent as $d$ increases. 
This is because of the $Z_d$ factor in the unparticle propagators, which is singular in the limit $d \to 2$. 
The $d=1$ peak in the CMS mono-$Z$ bound comes from the drop in the cross section at small $d$, as can be seen 
from Fig.~10 of \bib{CMS:2020ulv}.

Finally, we note that the upper limits presented in \fig{bound_U} apply for the absolute values of the couplings because 
we consider the four cases of unparticles separately. For the same reason, other results of this study are not sensitive to the 
sign of the unparticle-SM couplings.
\section{MUonE experiment}
\label{muone_exp}
The aim of the MUonE experiment is to measure precisely the shape of the differential cross section as it 
does not rely on the exact value of the luminosity \cite{Abbiendi:2677471}. 
From this shape, a template fit will be used to determine the value of $a_\mu^\text{had}$. 

\begin{figure}[h!]
  \centering
  \includegraphics[width=0.5\textwidth]{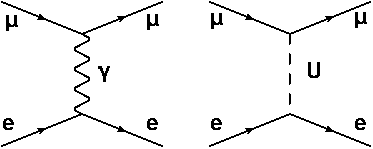}
  \caption{Feynman diagrams for the scattering process $e\mu \to e\mu$ with two contributions: the photon (left) and an 
  unparticle (right).}
  \label{MUonE_diag}
\end{figure}

The underlying scattering process reads
\begin{align}
e(p_e) + \mu(p_\mu) \to e(p'_e) + \mu(p'_\mu).
\label{eq:process}
\end{align}
The contributions from the photon and an unparticle can be visualized in \fig{MUonE_diag}.

At LO in the SM the unpolarized differential cross section reads
\begin{align}
\frac{d \sigma_\text{SM}}{dT}=\frac{\pi \alpha^2(t)}{(E_\mu^2-m_\mu^2)m_e^2 T^2}[2E_\mu m_e(E_\mu-T)-T(m_e^2+m_\mu^2-m_e T)],\label{SMcrosssection}
\end{align}
where 
\begin{align}
t = -2m_e T = (p_\mu - p'_\mu)^2, \quad T = E'_e - m_e \ge 0 ,\label{def_t_T}
\end{align}
with $p_\mu$ and $p'_\mu$ being the momentum of the initial-state and final-state muons, respectively. 
$E'_e$ is the energy of the final-state electron in the laboratory (Lab) frame. The variable $T$ is essentially 
$E'_e$ in practice.
The energy of the incoming muon in the laboratory frame is $E_\mu = 150\gev$ (using $160\gev$ as in \cite{Abbiendi:2022oks} does not 
change our conclusions). The center-of-mass energy is 
$\sqrt{s} = \sqrt{2E_\mu m_e + m_\mu^2 + m_e^2} \approx 0.4\gev$. Because of this low center-of-mass energy, the 
contribution from the $Z$ boson is negligible and has been removed from \eq{SMcrosssection}.

In the center of mass frame we have
\bea
t=-2|\vec{p}|^2 (1-\cos\theta), \quad |\vec{p}|^2 = \fr{s^2 + (m^2_\mu - m^2_e)^2 - 2s(m_\mu^2 + m_e^2)}{4s},
\label{kin_CMF}
\eea
where $\theta$ is the scattering angle. 
In the absence of kinematic cuts, the full range of $t$ is $[-4|\vec{p}|^2,0]$, which amounts to $[-0.143,0] \gev^2$ for $E_\mu = 150\gev$. The full range of $T$ is then $[0,139.818]$~GeV. 
In this paper, following \cite{Abbiendi:2016xup}, we impose a cut on the energy of the final-state electron $E'_e > 1\gev$ in the Lab frame. The range of $T$ then reads $[1,139.818]$~GeV. 

To calculate the $a_\mu^\text{had}$, it is conventional to parameterize $t$ in terms of the $x$ parameter as
\bea
t(x) = -\fr{m_\mu^2 x^2}{1-x}, \quad x \in [0,1).
\eea
We then have \cite{CarloniCalame:2015obs}
\bea
a_\mu^\text{had} = \fr{\alpha}{\pi}\int_0^{1} dx (1-x) \Delta\alpha_\text{had}[t(x)],
\label{amu_def}
\eea
where $\alpha = \alpha(0)$, $\Delta\alpha_\text{had}(t)$ is the hadronic contribution to the running of the coupling $\alpha$. 
The function $\Delta\alpha_\text{had}(t)$ is fitted from experimental data using \eq{SMcrosssection} with 
\bea
\alpha(t) = \fr{\alpha(0)}{1-\Delta\alpha(t)}, \quad \Delta\alpha(t) = \Delta\alpha_\text{had}(t) + \Delta\alpha_\text{lep}(t), 
\label{alpha_running}
\eea
where smaller contributions to the $\Delta\alpha(t)$ from the top quark and the $W^\pm$ bosons have been neglected. 
For $\Delta\alpha_\text{lep}(t)$, the precise prediction of the SM is used \cite{Abbiendi:2677471}.  
In this work, the values of $\Delta\alpha_\text{had}(t)$ and $\Delta\alpha_\text{lep}(t)$ are obtained using the 
package {\tt alphaQEDc19} \cite{Jegerlehner:2019lxt}.
\section{Unparticle effects at the MUonE experiment}
\label{unparticle_MUonE}
Switching on the unparticles (the right diagram of \fig{MUonE_diag}), the MUonE differential cross section then becomes, 
using the Feynman rules given in \tab{Feynman_rules},
\begin{align}
\frac{d \sigma_\text{new}}{dT}&=\frac{1}{128\pi m_e (E_\mu^2-m_\mu^2)} |\mathcal{M}_X +\mathcal{M}_\gamma|^2\crn
&=\frac{1}{128\pi m_e (E_\mu^2-m_\mu^2)} \left(\fr{16\pi^2 \alpha^2(t)}{t^2} \text{Tr}(\gamma\gamma) \right. \crn
&+ \left. \fr{8\pi \alpha(t)}{t} \fr{\lambda_X^2 Z_d |t-\mu^2|^{d-2}}{(M_Z^2)^{d-1}} \text{Tr}(\gamma X) 
+ \fr{\lambda_X^4 Z^2_d |t-\mu^2|^{2d-4}}{(M_Z^2)^{2d-2}} \text{Tr}(XX)
\right),\label{diff_XS_SMU}
\end{align}
where the results for $\text{Tr}(\gamma\gamma)$, $\text{Tr}(\gamma X)$, and $\text{Tr}(XX)$ are provided in \tab{traces}.  
The relative sign between the photon and the unparticle amplitudes 
is important as it affects the interference term. 
\begin{table}[h]
\begin{bigcenter}
\setlength\tabcolsep{0.05cm}
\begin{tabular}{|c|c|c|c|c|c|}
\hline
$Y$       & $\gamma$   & S & P & V & A \\ \hline
$\bar{\ell}\ell Y$ vertex & $ie\gamma_\mu$ & $i\lambda_S/M_Z^{d-1}$ & $-\lambda_P \gamma_5/M_Z^{d-1}$ & $i\lambda_V \gamma_\mu/M_Z^{d-1}$ & $i\lambda_A \gamma_\mu\gamma_5/M_Z^{d-1}$ 
               \\\hline
Propagator                & $-ig_{\mu\nu}/t$ & $iZ_{d}|t-\mu^2|^{d-2}$ & $iZ_{d}|t-\mu^2|^{d-2}$ 
& $-iZ_{d}|t-\mu^2|^{d-2}p_{\mu\nu}$ 
& $-iZ_{d}|t-\mu^2|^{d-2}p_{\mu\nu}$ \\
\hline               
\end{tabular}
\caption{Feynman rules for $e \mu \to e \mu$ scattering, where we have used the notation $p_{\mu\nu}=g_{\mu\nu}-ak_\mu k_\nu/t$ for the vector unparticles.}
\label{Feynman_rules}
\end{bigcenter}
\end{table}
\begin{table}[h]
\begin{center}
\begin{tabular}{|c|c|c|c|c|}
\hline
               & Type & $A_0$                                & $A_1$                                       & $A_2$       \\ \hline
$\text{Tr}(\gamma \gamma) $& &$64 E_\mu^2 m_e^2$             & $-64 E_\mu m_e^2-32 m_e^3-32 m_e m_\mu^2$ & $32m_e^2$ 
               \\\hline
$\text{Tr}(\gamma X)$ & S      & $-64 E_\mu m_e^2 m_\mu$             & $32 m_e^2 m_\mu $                        & $0$           \\ \cline{2-5} 
               & P      & $0$                                    & $0$                                         & $0$           \\ \cline{2-5} 
               & V      & $64 E_\mu^2 m_e^2$                       & $-64 E_\mu m_e^2-32 m_e^3-32 m_e m_\mu^2$ & $32m_e^2$   \\ 
\cline{2-5} 
               & A      & $0$  & $64 E_\mu m_e^2$ & $-32m_e^2$   \\ 
\hline
$\text{Tr}(XX)$       & S      & $64 m_e^2 m^2_\mu$                       & $32m_e^3+32 m_e m_\mu^2$                   & $16 m_e^2$  \\ \cline{2-5} 
               & P      & $0$                                    & $0$                                         & $16m_e^2$   \\ \cline{2-5} 
               & V      & $64 E_\mu^2 m_e^2$                       & $-64 E_\mu m_e^2-32 m_e^3-32 m_e m_\mu^2$ & $32m_e^2$   \\ 
\cline{2-5} 
               & A      & $64 E_\mu^2 m_e^2+64bm_e^2m_\mu^2$  & $-64 E_\mu m_e^2+32 m_e^3+32 m_e m_\mu^2$ & $32m_e^2$   \\ 
\hline
\end{tabular}
\caption{The coefficients $A_i$ of the traces, written as $A_0 + A_1 T + A_2 T^2$. For the $\text{Tr}(XX)$ 
of the axial-vector case, we have introduced an auxiliary parameter $b=2-2a-a^2$ in the coeffcient $A_0$.}
\label{traces}
\end{center}
\end{table}

In order to see the differences between different types of unparticles, we choose the following benchmark point
\begin{align}
\text{P0}:\quad d=1.1,\quad \lambda_i=0.02, \quad \mu=1\gev, 
\end{align}
where $i=S,P,V,A$. This point $\text{P0}$ satisfies all the constraints presented in \sect{sect:constraint}. 
From \fig{bound_U} one sees that $\text{P0}$ is at the edge of allowed region of the (axial-)vector couplings while there is still 
a good distance from it to the nearest bound of the (pseudo-)scalar couplings.
\begin{figure}[h!]
  \centering
  \begin{tabular}{cc}
  \includegraphics[width=0.49\textwidth]{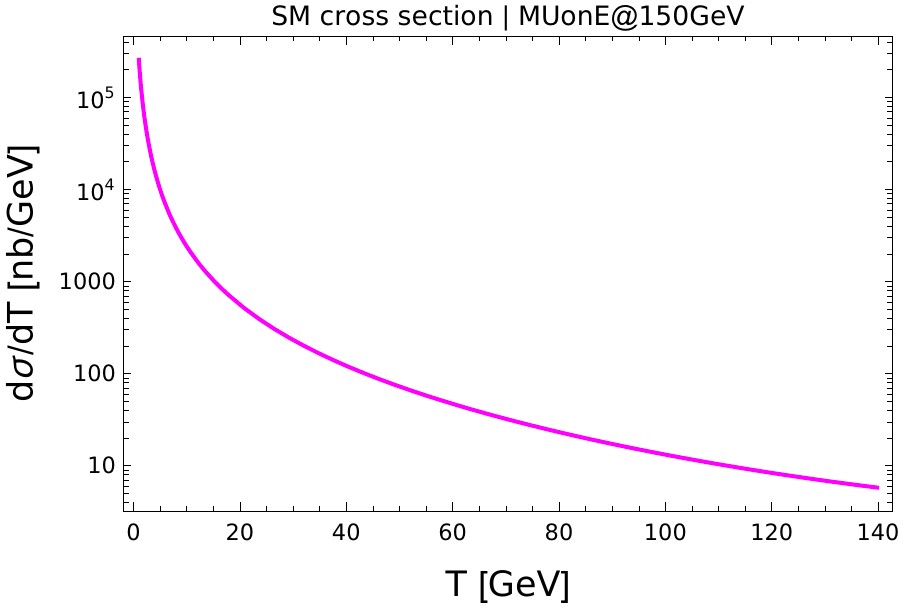} 
  \includegraphics[width=0.49\textwidth]{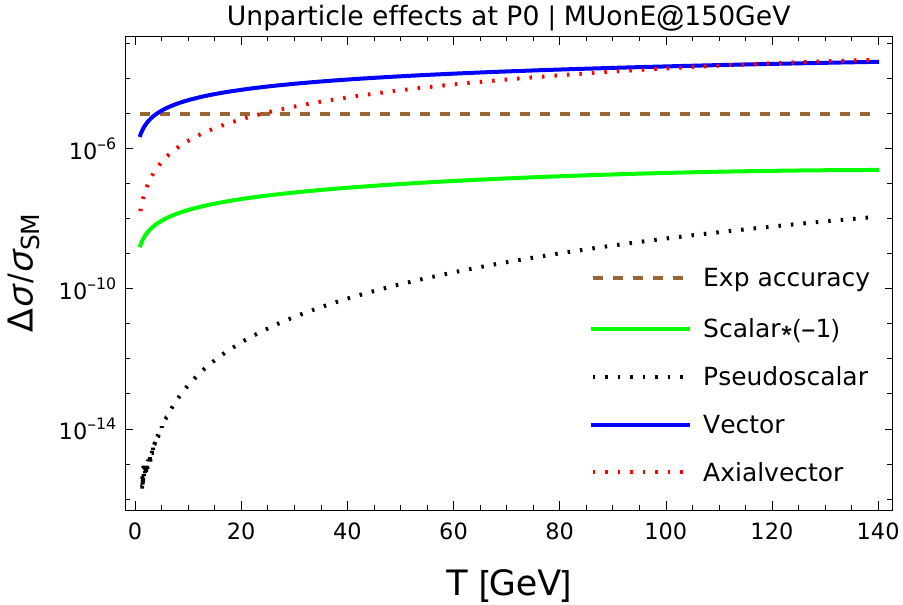}\\ 
  \end{tabular}
  \caption{Left: Differential cross section of the SM. Right: Various unparticle effects calculated at the parameter point $\text{P0}$ relative to the SM values. The MUonE systematic accuracy level of $10$ ppm is indicated by the dashed brown line.}
  \label{fig:SM_U_P0}
\end{figure}

In \fig{fig:SM_U_P0} we present the SM differential cross section $d\sigma/dT$ (left) and the four unparticle corrections to this distribution (right) calculated at the benchmark point $\text{P0}$ with respect to the SM values. The MUonE systematic accuracy level of $10$ ppm (dashed brown) is plotted to see whether unparticle effects can be observed by the detector. The results are interesting. We see that, with the same coupling strength, the vector unparticle effect (solid blue) is largest, then come the axial-vector (dotted red), scalar (solid green), pseudo-scalar (dotted black) in the order of decreasing magnitude. Notice that all unparticle effects are positive, except for the scalar case where we have switched its sign for the logarithmic-scale plotting. This is due to the destructive interference between the scalar unparticle and the photon amplitude.
The vector (axial-vector) unparticle effect is greater than the marked detector accuracy level when $T>3\gev$ ($24\gev$), indicating that the MUonE experiment may be able to explore the region around the point $\text{P0}$ in the 
parameter space. On the other hand, the (pseudo-)scalar unparticle curves are well below, more than one order of magnitude, the accuracy level. It is therefore impossible to detect the scalar unparticles around the $\text{P0}$ at the MUonE experiment. Indeed, we will later show that, by using a sensitivity threshold based on the $\chi^2$ defined in \eq{chi2_def}, it is impossible to observe (pseudo-)scalar unparticle effects in the entire parameter space region allowed by the LEP2 mono-photon bound.  

\begin{figure}[h!]
  \centering
  \begin{tabular}{cc}
  \includegraphics[width=0.49\textwidth]{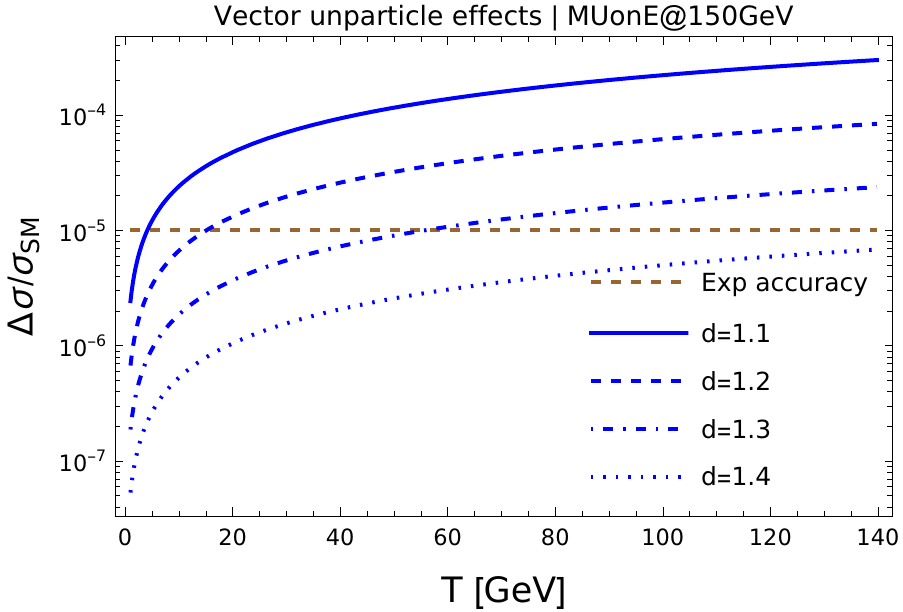} 
  \includegraphics[width=0.49\textwidth]{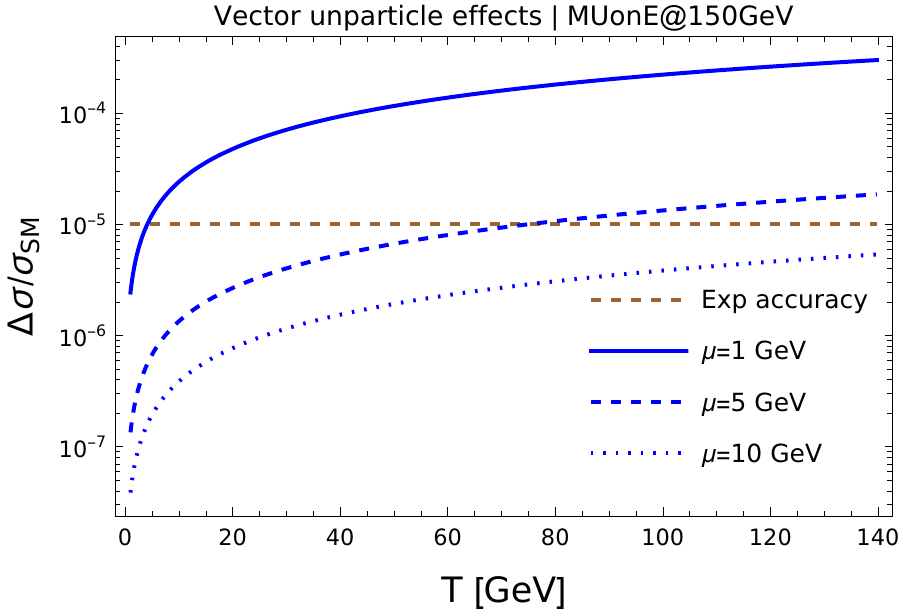}\\ 
  \end{tabular}
  \caption{Similar to the right plot in \fig{fig:SM_U_P0} but we plot here only the case of vector unparticle with 
  various values of the scaling dimension $d$ (left) and of the energy scale $\mu$ (right). The other parameters are kept at the $\text{P0}$ benchmark values.}
  \label{fig:V_P0_d_mu}
\end{figure}
We then vary the scaling dimension $d$ (\fig{fig:V_P0_d_mu} left) and the breaking-scale-invariance scale $\mu$ (\fig{fig:V_P0_d_mu} right) to see how the unparticle corrections depend on them. Since these dependences are the same for all unparticle types we plot only the vector case as a representative. As expected, the correction decreases monotonically with increasing $d$ or increasing $\mu$. 
With the other parameters being kept fixed at the $\text{P0}$ values, we see that the vector unparticle effect drops below the 
accuracy threshold for $d=1.4$ or $\mu=10\gev$. 

\begin{figure}[h!]
  \centering
  \includegraphics[width=0.49\textwidth]{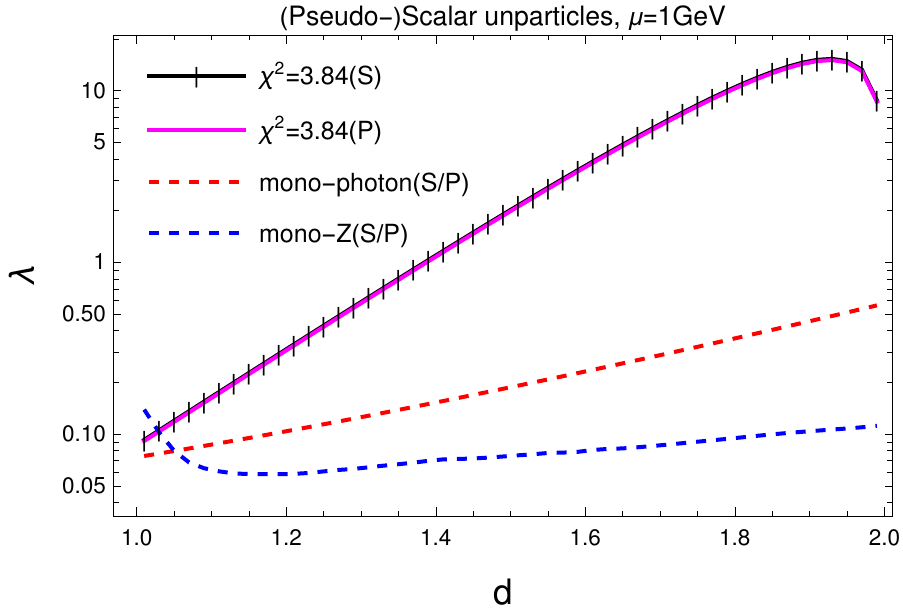}
  \includegraphics[width=0.49\textwidth]{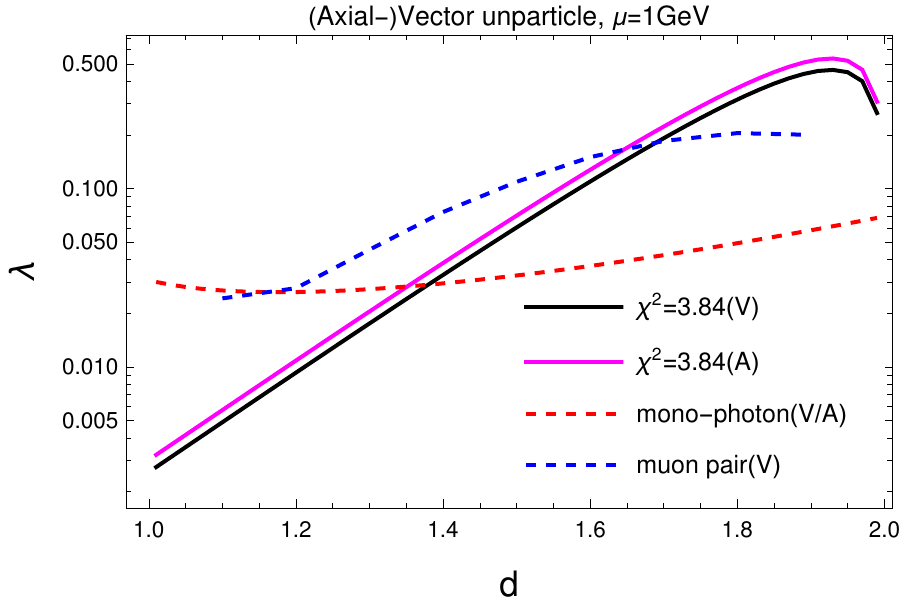}\\
  \includegraphics[width=0.49\textwidth]{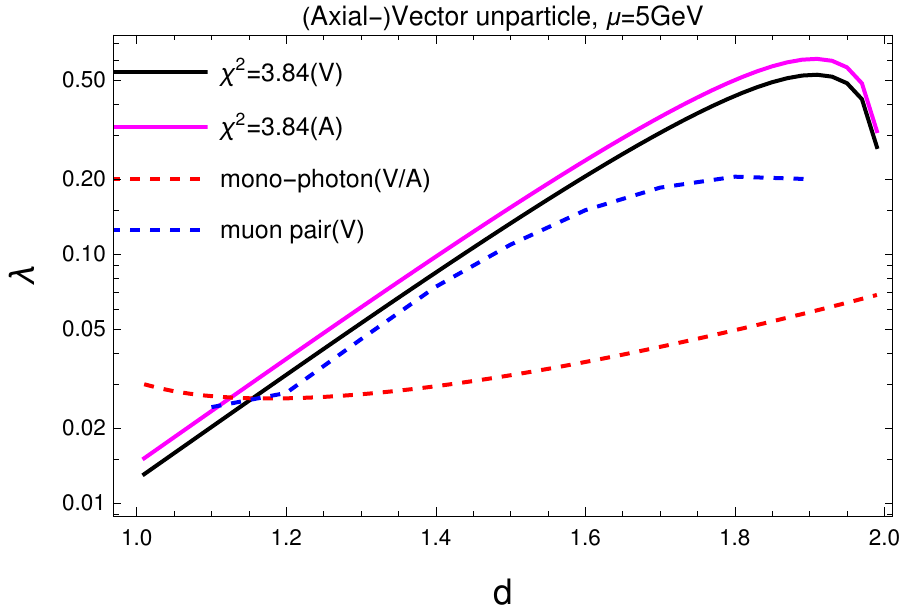}
  \includegraphics[width=0.49\textwidth]{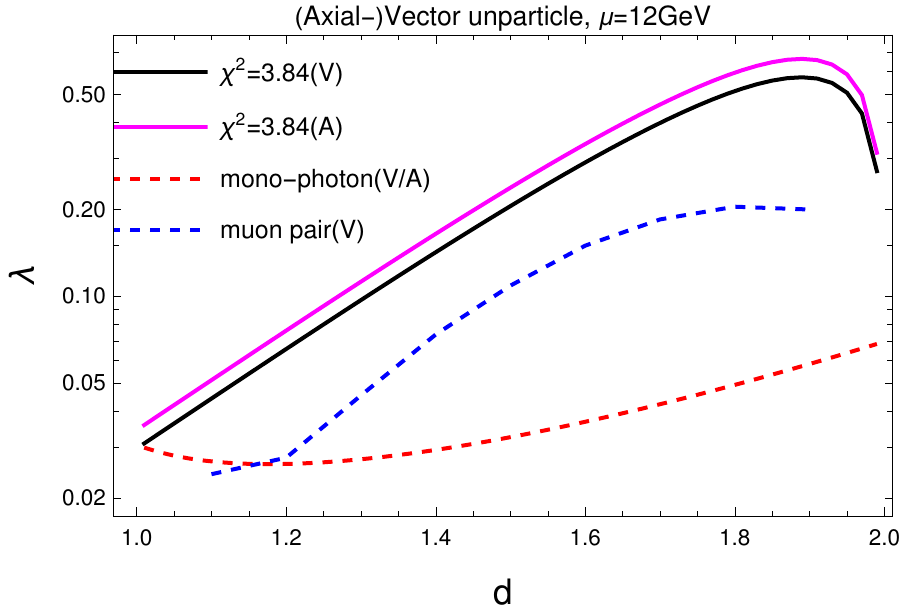}\\
  \caption{Sensitivity curves of the MUonE experiment on the unparticle-SM coupling $\lambda$ 
  for the cases of scalar (S), pseudo-scalar (P), vector (V), and axial-vector (A) unparticles. 
  The $95\%$ CL upper limits from the mono-photon, mono-$Z$, and muon-pair productions are also plotted. 
  The muon-pair bound for the pseudo-scalar case is not plotted as it is irrelevant.}
  \label{sens_curves}
\end{figure}
We now explore the region of $1.01 \le d \le 1.99$ and determine the sensitivity curves of the MUonE experiment on the unparticle-SM couplings $\lambda_i$. The sensitivity curves are defined at $\chi^2(d,\lambda_i,\mu)=3.84$, corresponding to a $95\%$ upper-limit CL. 
The $\chi^2$ is computed as \cite{Dev:2020drf}
\bea
\chi^2(d,\lambda,\mu) = \sum_{i=1}^{N_\text{bin}}\fr{\left(N_i^\text{new}(d,\lambda,\mu)-N_i^\text{SM}\right)^2}{\Delta_{i,\text{stat}}^2+\Delta_{i,\text{sys}}^2},\label{chi2_def}
\eea
where the statistical error $\Delta_{i,\text{stat}}=\sqrt{N_i^\text{new}}$ and the systematic error $\Delta_{i,\text{sys}}=10^{-5}N_i^\text{new}$. The number of events for each bin is calculated as
\bea
N_i = L\int_{T_i}^{T_i+\Delta T} \fr{d\sigma}{dT}(T) dT,
\eea
where $L=1.5\times 10^7$~$\text{nb}^{-1}$ is the integrated luminosity \cite{Abbiendi:2016xup} and the differential cross section 
is provided in \eq{SMcrosssection} and \eq{diff_XS_SMU}. 
We take $N_\text{bin}=30$ as in \cite{Abbiendi:2016xup}. 

Results for the scalar, pseudo-scalar, vector, and axial-vector unparticles are shown in \fig{sens_curves}. For the case 
of scalar unparticles, it is enough to show for the case of $\mu=1$ GeV as already at this minimum energy all 
parameter points satisfying the LEP mono-photon and CMS mono-$Z$ constraints lie below the sensitivity curves. 
Increasing the value of $\mu$ only pushes the sensitivity curves up. For the vector unparticles, results are 
presented for $\mu=1$, $5$, $12$ GeV. For $\mu > 12$ GeV the unparticle effects, when the couplings satisfy the current experimental 
constraints, are too small to be detected at the MUonE.
We therefore conclude that the MUonE experiment is insensitive to 
the (pseudo-)scalar unparticle effects for $\mu\ge 1$ GeV and sensitive to the (axial-)vector unparticles when 
$1\le \mu \lesssim 12$~GeV and $1<d\lesssim 1.4$. Sensitivities for the vector unparticle are slightly better than for the axial-vector case.

\begin{table}[ht]
\begin{center}
\begin{tabular}{|c|c|c|c|}
\hline
   &    SM    & Axial-vector  & Vector\\
\hline
$\text{P1}$ & $6903(29)\times 10^{-11}$ & $6957(29)\times 10^{-11}$ & $6986(29)\times 10^{-11}$ \\
Pull        & $0$        & $1.9$      & $2.8$  \\
\hline
$\text{P2}$ & $-$ & $6980(29)\times 10^{-11}$ & $7019(29)\times 10^{-11}$ \\
Pull        & $-$        & $2.6$      & $4.0$  \\
\hline
$\text{P3}$ & $-$ & $11073(29)\times 10^{-11}$ & $13250(29)\times 10^{-11}$ \\
Pull        & $-$        & $143$      & $218$  \\
\hline\hline
$\text{P4}$ & $-$ & $6954(29)\times 10^{-11}$ & $6979(29)\times 10^{-11}$ \\
Pull        & $-$        & $1.7$      & $2.6$  \\
\hline
$\text{P5}$ & $-$ & $6971(29)\times 10^{-11}$ & $7006(29)\times 10^{-11}$ \\
Pull        & $-$        & $2.3$      & $3.5$  \\
\hline
$\text{P6}$ & $-$ & $7091(29)\times 10^{-11}$ & $7186(29)\times 10^{-11}$ \\
Pull        & $-$        & $6.4$      & $9.7$  \\
\hline
\end{tabular}
\caption{Best-fit value of $a_\mu^\text{had}$ at the MUonE experiment in various scenarios: only SM, SM with axial-vector unparticle, SM with vector unparticle. The pull is defined as $(a_\mu^\text{had,new}-a_\mu^\text{had,SM})/\Delta_\text{SM}$, where $\Delta_\text{SM}$ is 
the error of the $a_\mu^\text{had,SM}$ under the assumption of no new physics.}
\label{fit_amu}
\end{center}
\end{table}

We now estimate the effects of the (axial-)vector unparticles on the determination of $a_\mu^\text{had}$. 
For this purpose, we choose the following benchmark points:  
\begin{align}
\text{P1}:&\quad d=1.01,\quad \lambda_i=0.0033, \quad \mu=1\gev,\\
\text{P2}:&\quad d=1.3,\quad \lambda_i=0.025, \quad \mu=1\gev,\\
\text{P3}:&\quad d=1.01,\quad \lambda_i=0.029, \quad \mu=1\gev,\\
\text{P4}:&\quad d=1.01,\quad \lambda_i=0.015, \quad \mu=5\gev,\\
\text{P5}:&\quad d=1.1,\quad \lambda_i=0.027, \quad \mu=5\gev,\\
\text{P6}:&\quad d=1.01,\quad \lambda_i=0.029, \quad \mu=5\gev,
\end{align}
where $i=V,A$. As can be seen from \fig{sens_curves}, these points lie above the sensitivity curves and below 
the upper-limit bound. For the case $\mu=1\gev$, the first three points $\text{P1}$, $\text{P2}$, $\text{P3}$ form a triangle region 
approximately covering the whole area of the parameter space which the MUonE is sensitive to and is still allowed by the current constraints. The points $\text{P4}$, $\text{P5}$, $\text{P6}$, for $\mu=5\gev$, are similarly chosen.  
The results are shown in \tab{fit_amu}.

These are obtained using a simplified $\chi^2$ fitting with one free parameter $k$, which 
enters the theoretical parametrization via the replacement $\Delta\alpha_\text{had}(t) \to k\Delta\alpha_\text{had}(t)$. 
The function $\Delta\alpha_\text{had}(t)$ is still calculated using the package {\tt alphaQEDc19}. 
The errors are computed from $\Delta\chi^2(k) =$ $\chi^2(k) - \chi^2_\text{min}$ $=1$. 

The best-fit result at the MUonE experiment with $1.5\times 10^7$ $\text{nb}^{-1}$ integrated luminosity for the SM-only case is $a_\mu^\text{had,SM}=6903(29)\times 10^{-11}$, where 
the mean value is given by the {\tt alphaQEDc19} program, agreeing within $1\sigma$ with the present 
world-average result of $6931(40)\times 10^{-11}$ provided in \bib{Workman:2022ynf}. The MUonE error of $0.4\%$, 
obtained using the $\chi^2$ defined in \eq{chi2_def}, is in accordance with the estimate of the experiment collaboration \cite{Abbiendi:2016xup}. 

The results show that if unparticles exist then the best-fit value of $a_\mu^\text{had}$ determined at the MUonE 
experiment will be larger than the one of the scenario without new physics. The increase (pull) is ranging 
from $1.7\sigma$ to $218\sigma$, and is higher for vector unparticle than for axial-vector case, being consistent 
with the results shown in \fig{fig:SM_U_P0} (right) and \fig{sens_curves}.
For the points lying exactly on the sensitivity curves the pull is $1.8\sigma$.   
\section{Conclusions}
\label{sect:conclusion}
Unparticle is a fascinating idea which offers us a novel form to describe the 
propagator of an intermediate stage. If this effect is visible at energy range around 
$0.5\gev$ then the newly proposed MUonE experiment, where the differential cross section 
of the elastic $e\mu \to e\mu$ scattering is measured at high accuracy, is a naturally 
good place to search for such effects. 

In this work we have considered the general case of unparticles with broken scale invariance, where the scale 
invariance of the unparticle system is broken at the energy $\mu$. In the limit of $\mu \to 0$ we recover the unbroken 
case originally proposed by Georgi. To avoid the severe limits placed on the couplings between unparticles and the SM fields from 
cosmology, astronomy, precise low-energy experiments one then requires $\mu \ge 1\gev$, which is sufficiently larger than 
the energy scales of the constraining processes. Under this condition, we have surveyed the relevant upper bounds on 
the unparticle-SM couplings from $e^+e^-\to \mu^+\mu^-$ experiments, forward-backward asymmetry data at LEP, 
mono-photon production at LEP2 and mono-$Z$ production at the LHC. As a by product, we have provided a new bound on the (pseudo-)scalar unparticle couplings using the mono-photon data at LEP2. 

Taking into account these constraints we then investigate the effects of unparticles at the MUonE experiment. 
We have considered separately four scenarios of unparticles: pseudo-scalar, scalar, axial-vector, and vector. 
This classification is based on the Lorentz structure of the unparticle-SM interaction terms. 
Given the same values for the input parameters (energy $\mu$, scaling dimension $d$, coupling $\lambda$), we found that the vector unparticle induces largest effects. Significantly smaller but still visible is the axial-vector case. The impact of 
(pseudo-)scalar unparticles are tiny and invisible. 

We have also studied the dependence of unparticle effects on the scaling dimension $d$ and on the energy scale $\mu$. 
As expected, the effect is largest when these parameters are small and decreases as they get larger. 
It was found that the MUonE experiment is sensitive to (axial-)vector unparticle effects when $1<d\lesssim 1.4$ and $1\le \mu \lesssim 12\gev$.

Finally, we have estimated the effect of the (axial-)vector unparticles on the $a_\mu^\text{had}$ measurement at the MUonE. 
Scanning the currently allowed parameter space, we found that unparticles can increase the best-fit value significantly. 
The MUonE experiment will then be able to provide hints of unparticles or new constraints on their couplings with the SM leptons if 
no significant deviation from the SM value is found. 

The above results are obtained for the MUonE with the muon-beam energy $E_\mu = 150\gev$. For $E_\mu = 160\gev$ as studied in \cite{Abbiendi:2022oks}, the c.m.s energy is slightly increased from $0.4055\gev$ to $0.4180\gev$. This change is too small to produce 
any visible differences in the results. Our conclusions are therefore unchanged.   
\acknowledgments

This research is funded by the Vietnam National Foundation for Science and
Technology Development (NAFOSTED) under grant number
103.01-2020.17.



\providecommand{\href}[2]{#2}\begingroup\raggedright\endgroup

\end{document}